\documentclass[final]{svjour3}
\usepackage{graphicx}
\usepackage{rotating}
\usepackage{amssymb}
\usepackage{mathptmx}
\usepackage{subfig}
\usepackage{diagbox}
\usepackage{color}
\usepackage{ulem}

\makeatletter
\journalname{Journal of Low Temperature Physics}

\begin{document}

\newcommand{\hdblarrow}{H\makebox[0.9ex][l]{$\downdownarrows$}-}
\title{Vibration characteristics of a continuously rotating superconducting magnetic bearing and potential influence to TES and SQUID}

\author{S.~Sugiyama\textsuperscript{a} \and
T.~Ghigna\textsuperscript{b} \and
Y.~Hoshino\textsuperscript{a} \and
N.~Katayama\textsuperscript{b} \and
S.~Katsuda\textsuperscript{a} \and
K.~Komatsu\textsuperscript{c} \and
T.~Matsumura\textsuperscript{b} \and
Y.~Sakurai\textsuperscript{c} \and
K.~Sato\textsuperscript{a} \and
R.~Takaku\textsuperscript{d} \and
M.~Tashiro\textsuperscript{a} \and
Y.~Terada\textsuperscript{a}}

\institute{
\textsuperscript{a} Saitama University, 255 Shimo-Okubo, Sakura-ku, Saitama, 338-8570, Japan
\textsuperscript{b} Kavli IPMU (WPI), UTIAS, The University of Tokyo, Kashiwa, Chiba 277-8583, Japan
\textsuperscript{c} Okayama University, 3-1-
1, Tsushimanaka, Kita-ku, Okayama City, Okayama, 700-8530, Japan
\textsuperscript{d} University of Tokyo, 7-3-1, Hongo, Bunkyo-ku, Tokyo 113-0033, Japan \\
\email{sugiyama@heal.phy.saitama-u.ac.jp}}


\maketitle

\begin{abstract}

We measured the vibration of a prototype superconducting magnetic bearing (SMB) operating at liquid nitrogen temperature. 
This prototype system was designed as a breadboard model for LiteBIRD low-frequency telescope (LFT) polarization modulator unit.
We set an upper limit of the vibration amplitude at 36~$\mu m$ at the rotational synchronous frequency.
During the rotation, the amplitude of the magnetic field produced varies. 
From this setup, we compute the static and AC amplitude of the magnetic fields produced by the SMB magnet at the location of the LFT focal plane as 0.24~G and $3\times10^{-5}$~G, respectively.
From the AC amplitude, we compute TES critical temperature variation of $7\times10^{-8}$~K and fractional change of the SQUID flux is $\delta \Phi/\Phi_0|_{ac}=3.1\times10^{-5}$. 
The mechanical vibration can be also estimated to be $3.6\times 10^{-2}$~N at the rotation mechanism location.



\end{abstract}

\vspace{-2mm}
\section{Introduction}
\vspace{-4mm}

The use of a continuously rotating half-wave plate (HWP) is a standard approach for cosmic microwave background (CMB) polarimetry to overcome detector $1/f$ noise.
Existing and upcoming CMB polarization experiments, including EBEX, POLARBAER, LSPE, Simons Observatory, employ HWPs with superconducting detector arrays\cite{ebex_hwp,polarbear_hwp,lspe_hwp,so_hwp}. 
Typically, the two sub-systems, HWP and detector array, are developed separately and integrated in the receiver's cryostat. 
Although a continuously rotating HWP is introduced to improve the polarimeter performance, a mechanically rotating element can introduce unexpected interference to a sensitive superconducting detector array. 
Such effects are usually identified and investigated only after integration resulting in delays of the scientific observations.
When it comes to a space mission, these delays during the integration phase can be critical. 
Thus, detailed planning for a campaign of detailed characterization and performance forecasting is essential.

One potential source of interference is due to vibrations of the continuously rotating HWP.
Mechanical vibrations propagating from the HWP to the detector system can affect detector noise and focal plane temperature stability. 
A cryogenically compatible HWP rotational mechanism employs a superconducting magnetic bearing (SMB). 
The rotor consists of a ring-shaped permanent magnet, hence its vibration can affect the detector system via magnetic coupling.  
Finally, mechanical motions other than the expected rotation can affect the incident radiation.

We study the vibration properties of a continuously rotating HWP supported by a SMB. 
There has been little experimental investigation of the vibration amplitude of a SMB rotor designed for CMB polarimetry.
Hull et al. reported measurements of the vibration amplitude of a SMB rotor~\cite{hull} with a rotor diameter of about 35~mm.
We report on the use of a Hall sensor to monitor the vibration via the change of magnetic field from the rotor magnet. 
This is a useful method to monitor the characteristic vibration frequency but not the physical displacement amplitude due to an extra complex calibration step~\cite{sakurai_magcal}.

LiteBIRD is the second L-class mission of ISAS/JAXA and its science goal is to probe the inflationary B-modes. 
The baseline configuration employs a continuously rotating HWP and transition-edge sensor (TES) bolometer arrays~\cite{lb_ptep_arxiv,litebird}.
The work reported here is part of developing an SMB-based polarization modulator for LiteBIRD.
In this paper we focus on characterizing the vibration properties of the SMB rotational mechanism and discuss their potential impact on the detector system. We also discuss future modeling of the impact of vibrations on the focal plane.

\begin{figure}[t]
\centering
\includegraphics[width=\textwidth]{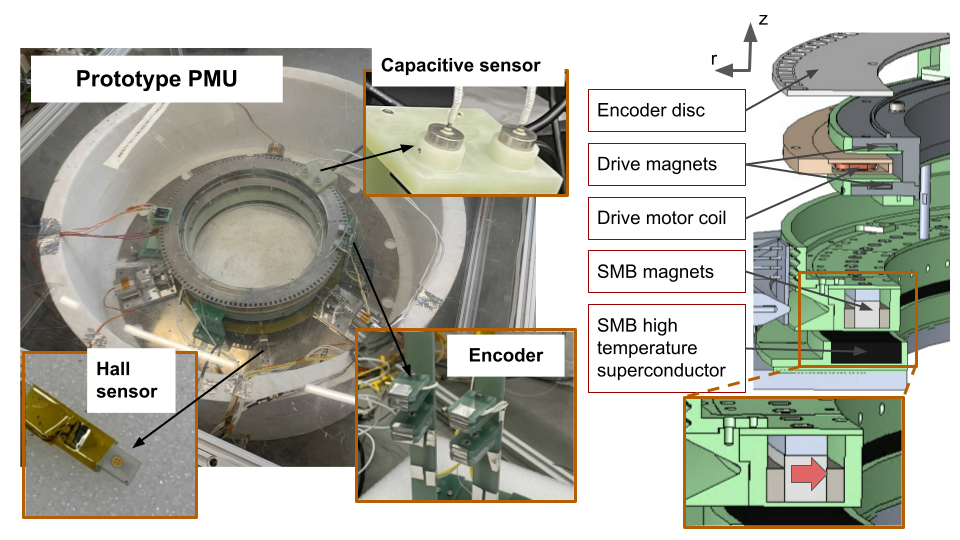}
\vspace{-4mm}
\caption{\footnotesize{A picture of the LiteBIRD LFT PMU breadboard model rotation mechanism (left) and a cross-section of the 3D CAD model (right). The arrow in the right bottom inset diagram indicates the orientation of the magnetization. }
\label{fig:PMU}}
\end{figure}

\vspace{-4mm}
\section{Experiment}
\vspace{-4mm}
\label{sec:experiment}
Figure~\ref{fig:PMU} shows the breadboard model of LiteBIRD LFT HWP rotational mechanism, which is scaled to about 0.7:1 with respect to the flight model. 
The rotor consists of a ring array of segmented magnets that is radially magnetized as shown in Figure~\ref{fig:PMU}.
The inner radius of the rotor magnet is 203~mm, the outer radius is 215~mm, and the thickness is 14~mm.
Due to its large diameter, the rotor magnet is segmented into 32 magnets. 
The stator consists of segmented YBCO high temperature superconductor array.
The 18 arc-shape bulk YBCO tiles form a ring shape with a diameter of 400~mm.
The rotor is driven by a contactless ac motor. 
The rotor magnets have alternating poles, and an array of coils is placed in the stator to drive the rotor. 
Detailed descriptions of each component can be found in Sakurai et al.~\cite{lft_pmu}.
The rotation of the rotor is monitored by two optical encoders \footnote{Hamamatsu, model number: L9337-01 and S2386-18L} used in combination with a 128-slots encoder disk mounted on the rotor. 
Axial displacement of the encoder disk is monitored by a capacitive sensor\footnote{Capacitec, model number: HPC-375E}. 
We also placed a Hall sensor\footnote{Lakeshore, model number: BHT-921} between the rotor and the stator. 

Before cooling the stator, the rotor is held by three cryogenic holder mechanisms at a distance of 5~mm from the stator. 
In order to test the system we submerge the stator superconductors in liquid nitrogen (LN2). 
Once the superconductors are thermalized to LN2 temperature, the cryogenic holder mechanisms release the rotor. 
Due to flux pinning the rotor is free to levitate and rotate in azimuth. The liquid nitrogen level is kept below the level of the rotor to reduce friction.
The rotor is driven by an ac motor at a constant rotational frequency of 1~Hz for 10~minutes. 
We limit the time duration of the data acquisition to maintain nearly the same LN2 level.
The data from the Hall and capacitive sensors are recorded simultaneously to correlate them. 
The capacitive sensor is calibrated separately prior to the measurements. 
Thus, we apply a calibration constant to convert the output voltage from the capacitive sensor to the rotor displacement in units of length.

\vspace{-6mm}
\section{Results}
\vspace{-4mm}
\label{sec:results}

\begin{figure}[htp]
\begin{tabular}{c}
\begin{minipage}[t]{0.5\hsize}
\includegraphics[width=\textwidth]{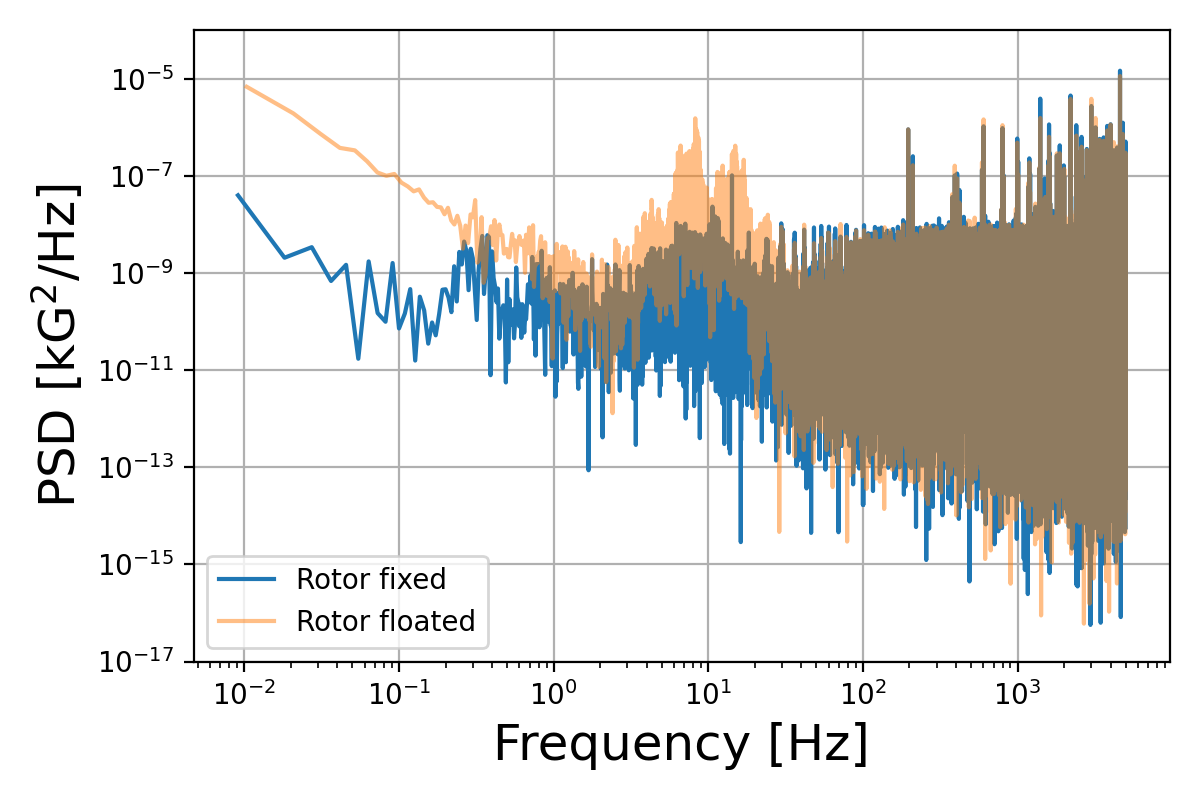}
\end{minipage}
\begin{minipage}[t]{0.5\hsize}
\includegraphics[width=1.0\textwidth]{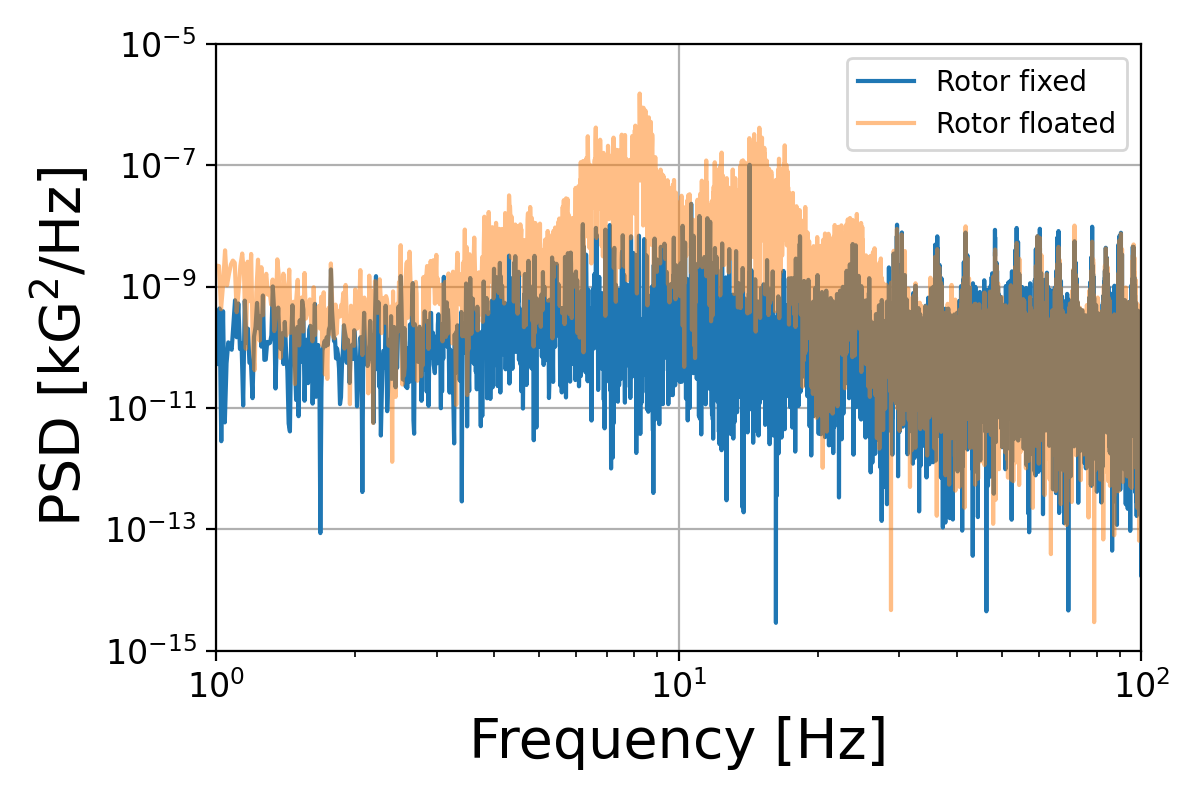}
\end{minipage}
\end{tabular}
\vspace{-4mm}
\caption{
\footnotesize{Left: the PSD of the magnetic field measured by the Hall sensor when the PMU rotor is gripped (blue) and levitated without rotation (orange). 
Right: Zoomed-in plot from 1\,Hz to 100\,Hz.}
\label{fig:Hall_close_open}}
\end{figure}

\begin{figure}[t]
\begin{tabular}{c}
\begin{minipage}[t]{0.5\hsize}
\includegraphics[width=\textwidth]{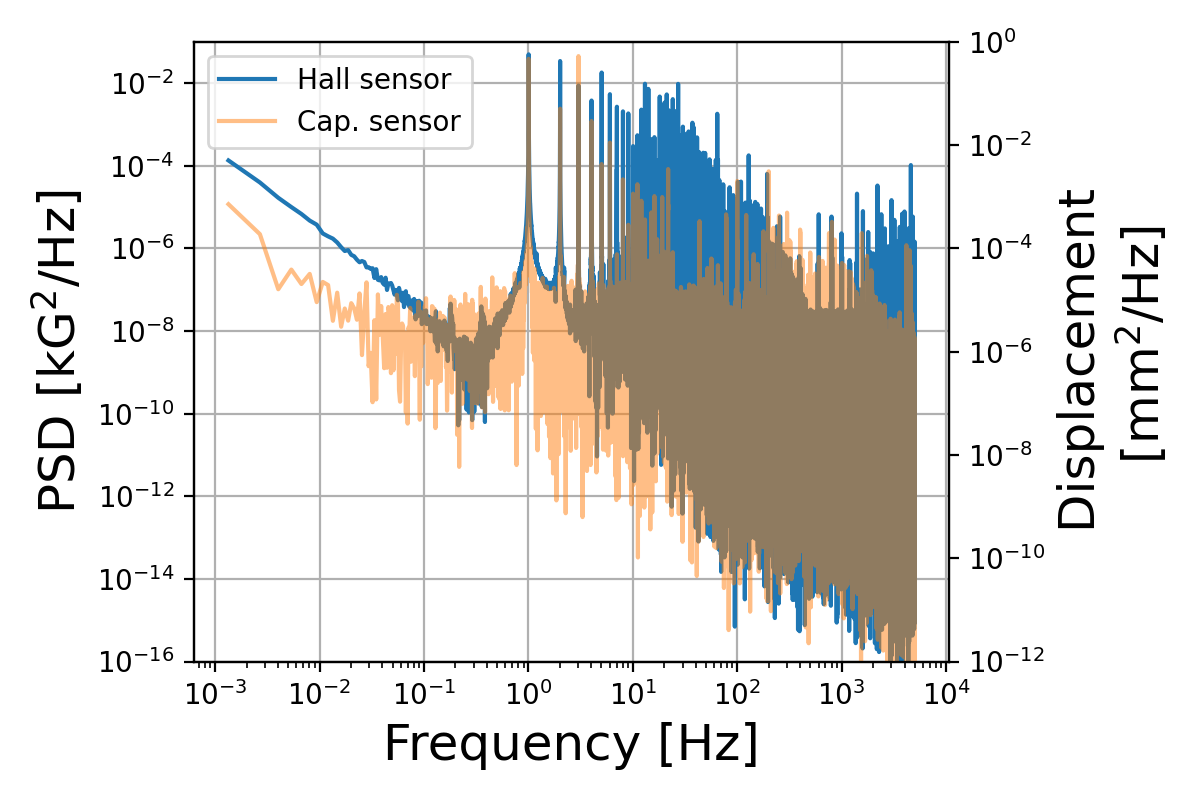}
\end{minipage}
\begin{minipage}[t]{0.5\hsize}
\includegraphics[width=1.0\textwidth]{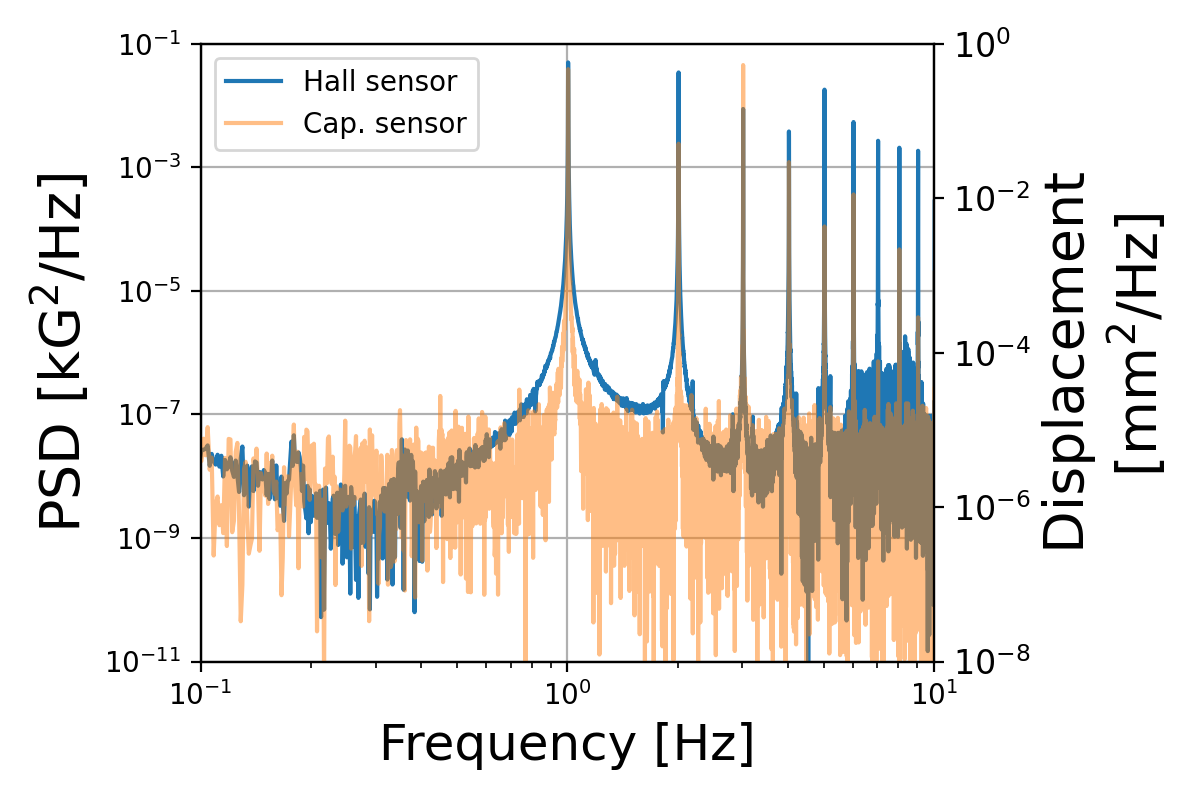}
\end{minipage}
\end{tabular}
\vspace{-4mm}
\caption{\footnotesize{Left: PSD of the Hall sensor output (blue) and the capacitive sensor output (orange) during rotation of the PMU rotor at 1~Hz.
Right: zoomed-in plot from 0.1\,Hz to 10\,Hz of the left panel plot to highlight the most relevant peaks. }
\label{fig:Hall_cap_stable}}
\end{figure}

\begin{figure}[t]
\centering
\includegraphics[width=\textwidth]{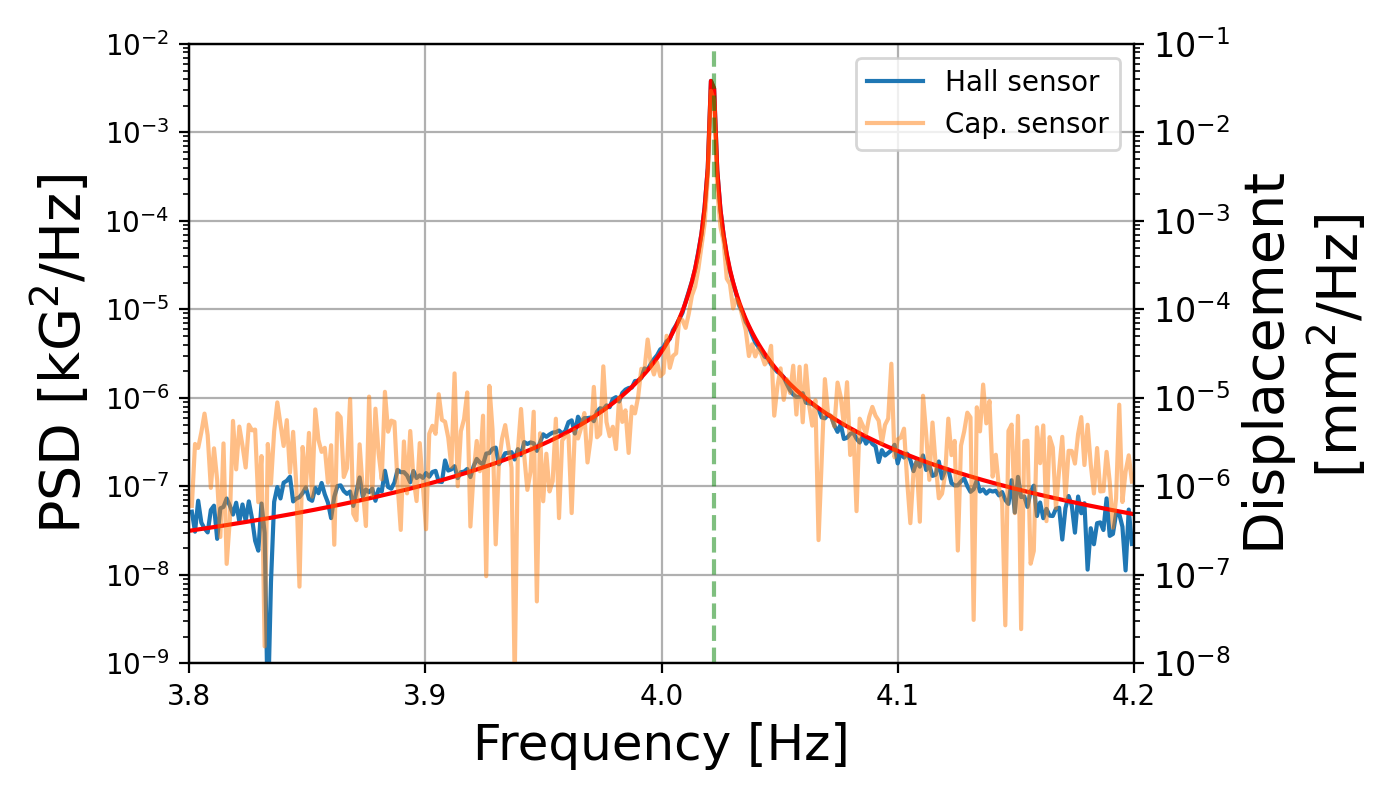}
\vspace{-5mm}
\caption{\footnotesize{Zoomed-in plot from 3.8\,Hz to 4.2\,Hz in Fig.~\ref{fig:Hall_cap_stable} shows around the fourth harmonic peak of the rotational frequency. We fit the magnetic field data (labeled Hall sensor) using
a power of forced oscillation with an attenuation, $\frac{F_0^2}{4 \pi^2 ((f - f_0)^2 + (2 \gamma f_0))^2}$ (red curve) with three fit parameters: an amplitude of forced oscillation as $2.5 \times 10^{-4}$, a center frequency of 4.02~Hz (green dash line), and an attenuation coefficient of $2.7 \times 10^{-9}$. We used the Hall sensor instead of the capacitive sensor to fit because the Hall sensor has a better sensitivity. We assume the magnetic field and the displacement is proportional to each other within the displacement less than 1~mm. }
\label{fig:psd_4Hz}}
\end{figure}

Fig.~\ref{fig:Hall_close_open} shows the power spectrum density~(PSD) of the magnetic field measured by the Hall sensor for two cases: (1) rotor gripped by the cryogenic holders, and (2) rotor levitated freely without rotation.
We can identify two broad peaks at around 8~Hz and 15~Hz and also minor peaks below and above the two peaks present only when the rotor is levitated.
These peaks correspond to natural frequencies in the radial and axial displacement and possibly other modes due to the nature of the segmented magnet array and YBCO array.
The main two peaks are consistent with previously reported peak locations~\cite{h_v_vibration}.
We also observe a series of peaks above 20~Hz in both gripped and levitating cases.
We attribute these to spurious effects due to the evaporation of LN2 in the vicinity of the Hall sensor, and thus we discard these peaks in both cases. 
Furthermore, the Hall sensor is read by a Lakeshore Gauss meter which filters above 400~Hz, and thus we do not investigate the cause of higher frequency peaks beyond this cut-off frequency.
We rely on the Hall sensor because the sensitivity of the capacitive sensor is not appropriate to detect nearly zero displacement during levitation.

Fig.~\ref{fig:Hall_cap_stable} show PSDs of the Hall and capacitive sensors when the rotor is spinning at 1 Hz.
Both PSDs show distinctive peaks at the rotational frequency and its harmonics.
The most prominent peak in the capacitive sensor PSD is at 1~Hz. 
From this measurement we find a corresponding physical displacement amplitude at 1~Hz of 36\,$\mathrm{\mu m}$. 
This value corresponds to a change in physical distance between the capacitive sensor and the encoder disk.
In principle, we may be measuring the deformation of the encoder disk in the axial direction even though the SMB rotor rotates with a perfect stiffness. In this paper, we assume that the encoder disk is perfectly flat and interpret the measured displacement as the amplitude of the axial rotor vibration.
Thus, we treat this number as an upper limit of the axial vibration amplitude. 

Figure~\ref{fig:psd_4Hz} shows the peak profile of the fourth harmonic. 
The shape is consistent with Lorentzian, i.e. a power of forced oscillation with an attenuation.
The corresponding width is parameterized by a Q-value equal to $3\times10^7$.  

\vspace{-4mm}
\section{Discussion}
\label{sec:discuss}
\vspace{-4mm}
In this section we discuss the potential effects of the HWP vibration on the TES detectors on the focal plane.
We use the LiteBIRD LFT configuration to define the relative position between the HWP and the focal plane~\cite{sekimoto}.
We also assume the amplitude of the vibration to have a maximum amplitude of 36~$\mu$m. 
We use this value for all vibration modes which we discuss in the following sections.




\begin{figure}[tbp]
\begin{tabular}{c}
\includegraphics[width=\textwidth]{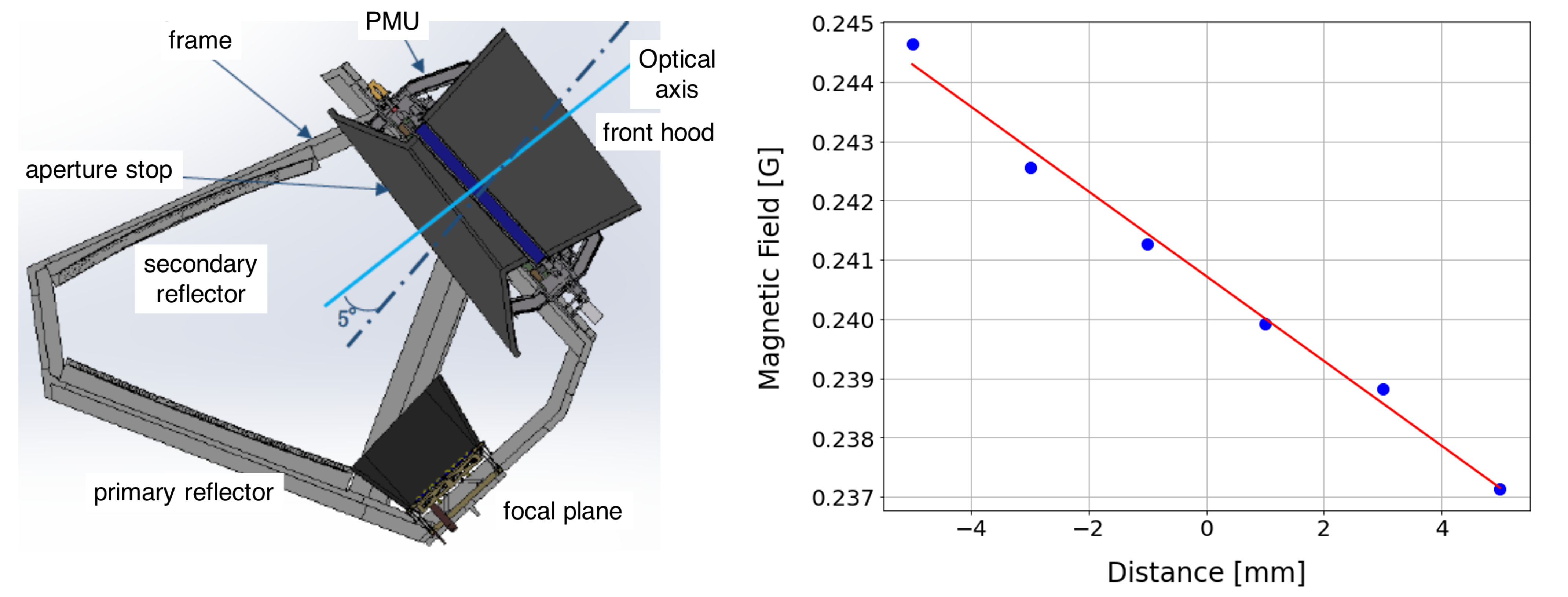}
\end{tabular}
\vspace{-4mm}
\caption{\footnotesize{Left: Cross-section view of the LiteBIRD LFT showing the relative positions of the PMU and the focal plane~\cite{sekimoto}.
Right: The magnitude of the magnetic field of the rotor at the center of the focal plane as a function of rotor displacement along the optical axis. We vary the ring magnet axially along the optical axis, and map the magnitude of the magnetic field.}
\label{fig:LFT_geo}}
\end{figure}

\vspace{-4mm}
\subsection{Varying magnetic field}
\vspace{-4mm}
We estimate the magnetic field at the focal plane due to a presence of the SMB rotor magnet at the aperture stop, dc field, and a displacement of the rotor magnet, ac field,  by using a commercially available finite element magnetic field simulation software, JMAG~\cite{jmag}.
We find the dc and ac variations of the magnetic field in the focal plane of 0.24~G and $3 \times 10^{-5}$~G in amplitude, respectively.
Detailed characterizations of the magnetic field susceptibility for LiteBIRD TES bolometers are in progress.
Therefore, we use literature values for the magnetic field susceptibility of ground-base experiments TESes.
Vavagiakis et al. reported a quadratic relationship between the critical temperature, $T_c$, and the applied magnetic field, $T_c=w B_{ext}^2 $~\cite{vavagiakis}. 
The highest sensitivity coefficient found in this study was measured for MoCu based TES designed for ACTpol, and corresponds to $w=4.6\times10^{-3}$~K/G$^2$. 
Due to the small applied magnetic field, we expand this quadratic form by the Taylor expansion as
$\Delta T_c \sim \frac{d T_c}{d B_{ext}} \Bigr|_{B_{ext}} \Delta B_{ext} = 2 w B_{ext} \Delta B_{ext}.$ 
This approach was taken in past studies~\cite{toki_phd,ghigna_ltd}.
As a result, we find 
$\Delta T_c = 7\times10^{-8}$~K.

The TES array is typically read by a Superconducting QUantum Interference Device (SQUID) that is sensitive to magnetic fields. 
LiteBIRD plans to use a frequency domain digital multiplexing (DfMUX) readout system~\cite{lb_ptep_arxiv,ghigna_ltd}. 
A similar system has been employed in past and ongoing CMB experiments~\cite{ebex_hwp,sptpol}.
The susceptibility of SQUIDs to external magnetic fields was investigated by Luker with a magnetic shield~\cite{mlueker_thesis}.
We assume a similar magnetic shielding scheme. 
The reported relationship between the fractional flux change and the external ac magnetic field $B_{ext}$ is $\delta \Phi/\Phi_0 = 0.06 + 1.03 B_{ext}$.
This results in a fractional flux $\delta \Phi/\Phi_0|_{ac} = 3.1\times10^{-5}$. 
This test was carried out with a non-gradiometric base SQUID fabricated at NIST~\cite{privatecommunication}. 
The new-generation SQUIDs are gradiometric and should be less sensitive to external magnetic fields.
Therefore, we report our result as an upper limit. 
Propagating the expected detector temperature fluctuations and SQUID responsivity variations to expected systematic uncertainties is left for future analysis. 

\vspace{-4mm}
\subsection{Restoring force} 
\vspace{-4mm}
When the rotor vibrates, mechanical vibrations can propagate to the detector system. 
A detailed propagation mechanism in a cryogenically cooled structure is complex and addressing this is beyond the scope of this paper.
Nevertheless, we can identify a similar situation by comparing the rotor vibration to those from the mechanical cooler. 

If we assume the vibrating rotor as a point-like mass supported by a spring, we can simply calculate the force required by the spring to realize this motion as 
\begin{eqnarray}
\vspace{-4mm}
F = m\ddot{x} = m A (2 \pi f)^2 \sin{(2 \pi ft)}, 
\vspace{-4mm}
\end{eqnarray}

where $m$ is the rotor mass. 
Here $A$ is the vibration amplitude and $f$ is the vibration frequency.
If we assume $m=25$~kg, $A=36~\mu$m and $f=1$~Hz as representative values for this exercise, the resultant force is $3.6\times 10^{-2}$~N.
Sato et al. reported on the measured force produced by a space qualified 1K Joule Thomson cooler~\cite{joule_thomson}. 
The maximum value is found at the driving frequency of 40~Hz in the range $0.1\sim10$~N depending on the orientation and the cooler.
The rotor vibration force estimated above is a factor of one or more smaller.
Again, further assessment of implication of this value is left for future analysis.

\vspace{-4mm}
\section{Conclusions}
\vspace{-4mm}
As part of the development of the breadboard model of the SMB-based HWP rotation mechanism for LiteBIRD LFT, we have measured the displacement of the rotor by using a capacitive sensor. 
We identified rotation synchronous peaks in the axial displacement data during rotation tests at 1~Hz. 
If we assume that the largest peak at 1~Hz is purely due to axial vibrations, we obtain a corresponding amplitude of 36~$\mu$m. 
From simulations we found that this displacement can generate an AC magnetic field of $3\times 10^{-5}$~G amplitude at the focal plane.
This corresponds to a prospective shift of $T_c$ of $7\times10^{-8}$~K if we adopt a magnetic field susceptibility from existing ground-based TES. 
We also estimated the effect  on the SQUID to be $\delta \Phi/\Phi_0|_{ac} = 3.1\times10^{-5}$.
The mechanical vibration force due to this vibration is found to be $3.6\times 10^{-2}$~N, which is a factor of one or more smaller than that from the 1K-JT cooler at the driving frequency of 40~Hz. 
The actual impact on the TES and focal plane temperature stability is yet to be investigated. 
This work is a part of an investigation of potential interference between the continuously rotating HWP and TES arrays for LiteBIRD LFT. 
While we investigated the source of the vibration experimentally, further studies are required to fully forecast the impact as noise and systematic effects through TES and SQUID.

\vspace{-4mm}
\begin{acknowledgements}
We acknowledge the World Premier International Research Center Initiative (WPI), MEXT, Japan for support through Kavli IPMU. This work was supported by JSPS KAKENHI Grant Numbers JP17H01125, 19K14732, 18KK0083, and JSPS Core-to-Core Program, A.Advanced Research Networks.
\end{acknowledgements}

\noindent \small{\bf{Ethical statement}}
\small{Our study is not applicable to the compliance with Ethical Standards.}

\pagebreak

\end{document}